\documentclass{ws-procs975x65}%
\usepackage{amsmath}
\usepackage{amsfonts}
\usepackage{amssymb}
\usepackage{graphicx}%
\def\beq{\begin{equation}}
\def\eeq{\end{equation}}

\begin{document}

\title{Traversable Wormholes and Yukawa Potentials}
\author{Remo Garattini}
\address{Universit\`{a} degli Studi di Bergamo, \\
Dipartimento di Ingegneria e Scienze Applicate,\\
Viale Marconi,5 24044 Dalmine (Bergamo) ITALY\\
I.N.F.N. - sezione di Milano, Milan, Italy\\
$^*$E-mail:remo.garattini@unibg.it}

\begin{abstract}
Traversable Wormhole are amazing astrophysical objects predicted by General
Relativity which are able to connect remote region of space-time. Even if
their existence has not been proved yet they are object of continuous
investigation. From the theoretically point of view, to exist, traversable
wormholes need a special form of energy density termed \textquotedblleft%
\textit{exotic}\textquotedblright. Since this exotic source must be
concentrated on the throat of the wormhole, we discuss the implications of
assuming Yukawa-like profiles which could be realize such a configuration.

\end{abstract}

\bodymatter

\section{Introduction}

Traversable wormholes are solutions of the Einstein's Field Equations (EFE)
which have the property of connecting remote space-time
regions\cite{MT,Visser}. In Schwarzschild coordinates, the traversable
wormhole metric can be cast into the form%
\begin{equation}
ds^{2}=-\exp\left(  -2\phi\left(  r\right)  \right)  dt^{2}+\frac{dr^{2}%
}{1-\frac{b\left(  r\right)  }{r}}+r^{2}d\Omega^{2}.\label{metric}%
\end{equation}
where $\phi\left(  r\right)  $ is called the redshift function, while
$b\left(  r\right)  $ is called the shape function and where $d\Omega
^{2}=d\theta^{2}+\sin^{2}\theta d\phi^{2}$ is the line element of the unit
sphere. $\phi(r)$ and $b(r)$ are arbitrary functions of the radial coordinate
$r\in\left[  r_{0},+\infty\right)  $. A fundamental property of a wormhole is
that a flaring out condition of the throat, given by $(b-b^{\prime}r)/b^{2}%
>0$, must be satisfied as well as the request that $1-b(r)/r>0$. Furthermore,
at the throat $b(r_{0})=r_{0}$ and the condition $b^{\prime}(r_{0})<1$ is
imposed to have wormhole solutions. It is also fundamental that there are no
horizons present, which are identified as the surfaces with $e^{2\phi
}\rightarrow0$, so that $\phi(r)$ must be finite everywhere. The first step to
establish if a traversable wormhole exists is given by solving the EFE in an
orthonormal reference frame%
\begin{equation}
\rho\left(  r\right)  =\frac{1}{8\pi G}\frac{b^{\prime}}{r^{2}},\label{rho}%
\end{equation}%
\begin{equation}
p_{r}\left(  r\right)  =\frac{1}{8\pi G}\left[  \frac{2}{r}\left(
1-\frac{b\left(  r\right)  }{r}\right)  \phi^{\prime}-\frac{b}{r^{3}}\right]
,\label{pr}%
\end{equation}%
\begin{equation}
p_{t}\left(  r\right)  =\frac{1}{8\pi G}\left(  1-\frac{b\left(  r\right)
}{r}\right)  \left[  \phi^{\prime\prime}+\phi^{\prime}\left(  \phi^{\prime
}+\frac{1}{r}\right)  \right]  -\frac{b^{\prime}r-b}{2r^{2}}\left(
\phi^{\prime}+\frac{1}{r}\right)  ,\label{pt}%
\end{equation}
where $\rho\left(  r\right)  $ is the energy density, $p_{r}\left(  r\right)
$ is the radial pressure, and $p_{t}\left(  r\right)  $ is the lateral
pressure and using the conservation of the stress-energy tensor, in the same
orthonormal reference frame, one gets%
\begin{equation}
p_{r}^{\prime}=\frac{2}{r}\left(  p_{t}-p_{r}\right)  -\left(  \rho
+p_{r}\right)  \phi^{\prime}.
\end{equation}
One strategy to obtain solutions for the EFE is represented by imposing an
Equation of State (EoS) of the following forms:

\begin{arabiclist}
\item $p_{r}(r)=\omega\rho(r)$, with $\omega$ constant,

\item $p_{r}(r)=\omega\left(  r\right)  \rho(r)$, with $\omega$ function of
the radial coordinate $r$,

\item $p_{r}(r)=\omega\rho^{\gamma}(r)$, with $\gamma\in%
\mathbb{R}
$.
\end{arabiclist}

Of course, this list does not exhaust the possibilities of constraining the
relationship between $p_{r}(r)$ and $\rho(r)$. For instance, by imposing the
first EoS, one finds%
\begin{equation}
\phi^{\prime}=\frac{b\left(  r\right)  +\omega b\left(  r\right)  ^{\prime}%
r}{2r^{2}\left(  1-\frac{b\left(  r\right)  }{r}\right)  } \label{phi'w}%
\end{equation}
and if we also assume Zero Tidal Forces (ZTF), one gets%
\begin{equation}
\phi\left(  r\right)  =C\qquad\mathrm{and}\qquad b\left(  r\right)
=r_{0}\left(  \frac{r_{0}}{r}\right)  ^{\frac{1}{\omega}}, \label{b(r)0}%
\end{equation}
where the condition $b\left(  r_{0}\right)  =r_{0}$ has been used. The
parameter $\omega$ is restricted by the following conditions%
\begin{equation}
b^{\prime}\left(  r_{0}\right)  <1;\qquad\frac{b\left(  r\right)  }%
{r}\underset{r\rightarrow+\infty}{\rightarrow0}\qquad\Longrightarrow
\qquad\left\{
\begin{array}
[c]{c}%
\omega>0\\
\omega<-1
\end{array}
\right.
\end{equation}
and the metric $\left(  \ref{metric}\right)  $ assumes the particular simple
expression%
\begin{equation}
ds^{2}=-dt^{2}+\frac{dr^{2}}{1-\left(  \frac{r_{0}}{r}\right)  ^{\frac
{1}{\omega}+1}}+r^{2}d\Omega^{2}.
\end{equation}
It is interesting to note that when $\omega\rightarrow\infty$, one finds%
\begin{equation}
ds^{2}=-dt^{2}+\frac{dr^{2}}{1-\frac{r_{0}}{r}}+r^{2}d\Omega^{2}, \label{STW}%
\end{equation}
namely a traversable wormhole with zero energy density and ZTF. The
corresponding Stress-Energy Tensor (SET) is given by%
\begin{equation}
T_{\mu\nu}=\left(  \rho\left(  r\right)  ,p_{r}\left(  r\right)  ,p_{t}\left(
r\right)  ,p_{t}\left(  r\right)  \right)  =\frac{r_{0}}{2\kappa r^{3}}\left(
0,0,1,1\right)  ,
\end{equation}
that is it is formed by pure transverse pressure. To establish how much
negative energy density is necessary, it is useful the computation of the
\textquotedblleft volume integral quantifier\textquotedblright, which provides
information about the \textquotedblleft total amount\textquotedblright\ of
averaged null energy condition (ANEC) violating matter in the
space-time\cite{VKD}. This is defined by%
\begin{equation}
I_{V}=\int[\rho(r)+p_{r}(r)]dV
\end{equation}
and for the line element $\left(  \ref{metric}\right)  $, one can write%
\begin{equation}
I_{V}=\frac{1}{\kappa}\int_{r_{0}}^{+\infty}\left(  r-b\left(  r\right)
\right)  \left[  \ln\left(  \frac{e^{2\phi(r)}}{1-\frac{b\left(  r\right)
}{r}}\right)  \right]  ^{\prime}dr. \label{IV}%
\end{equation}
For instance, for the metric $\left(  \ref{STW}\right)  $, one finds%
\begin{equation}
I_{V}=-\frac{1}{\kappa}\int_{r_{0}}^{+\infty}\left(  r-r_{0}\right)  \left[
\ln\left(  1-\frac{r_{0}}{r}\right)  \right]  ^{\prime}dr=-\frac{r_{0}}%
{\kappa}\left[  \ln\left(  r\right)  \right]  _{r_{0}}^{+\infty}%
\rightarrow-\infty
\end{equation}
which means that an infinite amount of negative energy is necessary to build
such a wormhole. This was also confirmed in Ref.\cite{MT}. For this reason, we
are going to explore the possibilities offered by the second EoS. The third
EoS requires a careful analysis and it will not presented here. In particular,
we would like to consider Yukawa-type profiles for $b\left(  r\right)  $ and
$\omega\left(  r\right)  $. The hope is that such a profile concentrates more
the energy density to the wormhole throat and thus minimizes the usage of
exotic matter.

\section{The Inhomogeneous Equation of State and the Volume Integral
Quantifier}

\label{p1}When we apply the inhomogeneous EoS $p_{r}=\omega\left(  r\right)
\rho$ to find the corresponding shape function, one finds%
\begin{equation}
b(r)=r_{0}\,\exp\left[  -\int_{r_{0}}^{r}\,\frac{d\bar{r}}{\omega(\bar{r}%
)\bar{r}}\right]  \,. \label{b(r)}%
\end{equation}
The shape function $\left(  \ref{b(r)}\right)  $ is obtained by imposing
$\phi^{\prime}\left(  r\right)  =0$. Since we know the form of the redshift
function and of the shape function, the SET can also be easily computed
\begin{equation}
T_{\mu\nu}=\frac{r_{0}}{\kappa r^{3}}\left(  -\frac{1}{\omega\left(  r\right)
},-1,\frac{1}{2\omega\left(  r\right)  }+\frac{1}{2},\frac{1}{2\omega\left(
r\right)  }+\frac{1}{2}\right)  \exp\left[  -\int_{r_{0}}^{r}\,\frac{d\bar{r}%
}{\omega(\bar{r})\bar{r}}\right]  . \label{SET}%
\end{equation}
Note that the SET $\left(  \ref{SET}\right)  $ is traceless. We will examine
two specific choices for $\omega\left(  r\right)  $.

\subsection{Two examples for $\omega\left(  r\right)  $}

We are going to focus our attention on two examples. The first one is
represented by%
\begin{equation}
\omega\left(  r\right)  =\frac{1}{\mu r}\qquad\Longrightarrow\qquad
\omega\left(  r_{0}\right)  =\frac{1}{\mu r_{0}}, \label{omega}%
\end{equation}
leading to%
\begin{equation}
b(r)=r_{0}\exp\left[  -\mu\left(  r-r_{0}\right)  \right]  \qquad and\qquad
b^{\prime}(r)=-\mu r_{0}\exp\left[  -\mu\left(  r-r_{0}\right)  \right]
\end{equation}
satisfying therefore the flare-out condition since $b^{\prime}(r_{0})=-\mu
r_{0}$. The form of the metric $\left(  \ref{metric}\right)  $ therefore
becomes%
\begin{equation}
ds^{2}=-dt^{2}+\frac{dr^{2}}{1-\frac{r_{0}\exp\left[  -\mu\left(
r-r_{0}\right)  \right]  }{r}}+r^{2}d\Omega^{2}, \label{ds1}%
\end{equation}
which looks like a Yukawa profile, at least on the radial part. From the line
element $\left(  \ref{ds1}\right)  $, the SET is easily computed and assumes
the following form%
\begin{equation}
T_{\mu\nu}=\frac{r_{0}}{2\kappa{r}^{3}}\left[  diag\left(  -2\mu r,-2,\mu
r+1,\mu r+1\right)  \right]  \exp\left[  -\mu\left(  r-r_{0}\right)  \right]
\,. \label{SET1}%
\end{equation}
Note that the SET is also traceless as it should be, because it is a
particular case of the SET $\left(  \ref{SET}\right)  $. On the throat the
SET\ becomes%
\begin{equation}
T_{\mu\nu}=\frac{1}{2\kappa{r}_{0}^{2}}\left[  diag\left(  -2\mu r_{0},-2,\mu
r_{0}+1,\mu r_{0}+1\right)  \right]  \,
\end{equation}
and the following property is satisfied%
\begin{equation}
\lim_{\mu{{\rightarrow0}}}\lim_{r\rightarrow r_{0}}T_{\mu\nu}=\lim
_{r\rightarrow r_{0}}\lim_{\mu{{\rightarrow0}}}T_{\mu\nu}.
\end{equation}
On the other hand, we find%
\begin{equation}
\lim_{\mu{{\rightarrow\infty}}}\lim_{r\rightarrow r_{0}}T_{\mu\nu}\neq
\lim_{r\rightarrow r_{0}}\lim_{\mu{{\rightarrow\infty}}}T_{\mu\nu}.
\end{equation}
In particular, $\underset{\mu\rightarrow\infty}{\lim}$ $\underset{r-r_{0}%
}{\lim}T_{\mu\nu}$ is not defined, while $\left(  \ref{omega1}\right)  $%
\begin{equation}
\lim_{r\rightarrow r_{0}}\lim_{\mu{{\rightarrow\infty}}}T_{\mu\nu}=\left[
diag\left(  0,0,0,0\right)  \right]
\end{equation}
corresponding to the Minkowski space written in spherical coordinates. This
can be confirmed also by looking at the line element $\left(  \ref{ds1}%
\right)  $. The corresponding $I_{V}$ for the line element $\left(
\ref{metric}\right)  $ becomes%
\begin{gather}
I_{V}=\frac{1}{\kappa}\int_{r_{0}}^{+\infty}\left(  r-r_{0}e^{-\mu\left(
r-r_{0}\right)  }\right)  \left[  \ln\left(  1-\frac{r_{0}}{r}e^{-\mu\left(
r-r_{0}\right)  }\right)  \right]  ^{\prime}dr\nonumber\\
=\frac{r_{0}}{\kappa}\left(  1+e^{\mu r_{0}}\operatorname{Ei}_{1}\left(  \mu
r_{0}\right)  \right)  ,
\end{gather}
where $\operatorname{Ei}_{1}\left(  x\right)  $ is the exponential integral.
Since the integrand is finite, we can easily evaluate the behavior close to
the throat, whose result is%
\begin{gather}
I_{V}\simeq\frac{r_{0}}{\kappa}\,\left(  1-\exp\left(  \mu r_{0}\right)
\operatorname{Ei}_{1}\left(  \mu\left(  r_{0}+\varepsilon\right)  \right)
\,+\exp\left(  \mu r_{0}\right)  \operatorname{Ei}_{1}\left(  \mu
r_{0}\right)  -\exp\left(  -\mu\varepsilon\right)  \right)  \rightarrow
0\nonumber\\
\qquad\mathrm{when}\text{ }\varepsilon\rightarrow0,
\end{gather}
namely $I_{V}$ can be arbitrarily small. The second example we are going to
discuss is the following%
\begin{equation}
\omega\left(  r\right)  =\frac{\exp\left[  -\mu\left(  r-r_{0}\right)
\right]  }{\mu r}, \label{omega1}%
\end{equation}
which has the following properties%
\begin{equation}
\omega\left(  r\right)  \underset{r{{\rightarrow\infty}}}{\rightarrow}0\qquad
and\qquad\omega\left(  r_{0}\right)  =\frac{1}{\mu r_{0}}.
\end{equation}
Plugging $\left(  \ref{omega1}\right)  $ into the shape function $\left(
\ref{b(r)}\right)  $, one finds%
\begin{align}
b(r)  &  =r_{0}\exp\left[  1-\exp\left(  \mu\left(  r-r_{0}\right)  \right)
\right] \nonumber\\
b^{\prime}(r)  &  =-\mu r_{0}\,\exp\left[  \mu\,\left(  r-r_{0}\right)
+1-{\mathrm{\exp}}\left(  \mu\,\left(  r-r_{0}\right)  \right)  \right]  ,
\end{align}
satisfying the flare-out condition since%
\begin{equation}
b^{\prime}(r_{0})=-\mu r_{0}\,.
\end{equation}
Even in this case, the SET is easily computed and assumes the explicit form%
\begin{gather}
T_{\mu\nu}\nonumber\\
=\frac{r_{0}}{2\kappa r^{3}}\left[  diag\left(  -\frac{2}{\omega\left(
r\right)  },-2,\frac{1}{\omega\left(  r\right)  }+1,\frac{1}{\omega\left(
r\right)  }+1\right)  \right]  \exp\left[  1-\exp\left(  \mu\left(
r-r_{0}\right)  \right)  \right]  , \label{SET2}%
\end{gather}
where $\omega\left(  r\right)  $ is given by Eq.$\left(  \ref{omega1}\right)
$. On the throat, one gets%
\begin{equation}
T_{\mu\nu}=\frac{1}{2\kappa r_{0}^{2}}\left[  diag\left(  -2\mu r_{0},-2,\mu
r_{0}+1,\mu r_{0}+1\right)  \right]  \,
\end{equation}
and also in this case we obtain the same behavior of the SET $\left(
\ref{SET1}\right)  $. It is interesting to note that, on the throat, the
behavior of the SET in $\left(  \ref{SET1}\right)  $ and in $\left(
\ref{SET2}\right)  $ are equal. The corresponding $I_{V}$ for the line element
$\left(  \ref{metric}\right)  $ becomes%
\begin{equation}
I_{V}=-\frac{1}{\kappa}\int_{r_{0}}^{+\infty}\left(  r-r_{0}\exp\left[
1-\exp\left(  \mu\left(  r-r_{0}\right)  \right)  \right]  \right)  \left[
\ln\left(  1-\frac{r_{0}\exp\left[  1-\exp\left(  \mu\left(  r-r_{0}\right)
\right)  \right]  }{r}\right)  \right]  ^{\prime}dr=-\frac{1}{\kappa}%
\int_{r_{0}}^{+\infty}f\left(  r\right)  dr,
\end{equation}
where%
\begin{equation}
f\left(  r\right)  =\frac{r_{0}\exp\left[  1-\exp\left(  \mu\left(
r-r_{0}\right)  \right)  \right]  \left(  \exp\left(  \mu\left(
r-r_{0}\right)  \right)  \mu\,r+1\right)  }{r}.
\end{equation}
The exact evaluation of $I_{V}$ is quite complicated but, an estimate when
$r\gg r_{0}$ is possible to see if the integral is finite for large $r$.
Indeed, one finds that the integrand becomes%
\begin{align}
&  f\left(  r\right)  \underset{r{\rightarrow\infty}}{\simeq}{\frac{\mu
r_{0}\exp\left(  \mu\left(  r-r_{0}\right)  +1\right)  }{\exp\left(
\exp\left(  \mu\left(  r-r_{0}\right)  \right)  \right)  }}\nonumber\\
&  {\Longrightarrow I_{V}\simeq-}r_{0}{\exp\left(  1-\exp\left(  \mu\left(
r-r_{0}\right)  \right)  \right)  \rightarrow0\qquad\mathrm{when}}\text{
}r{\rightarrow\infty\qquad}%
\end{align}
\noindent and therefore $I_{V}$ will be finite close to infinity. On the other
hand, close to the throat, we can write%
\begin{equation}
I_{V}\simeq\frac{1}{\kappa}\int_{r_{0}}^{r_{0}+\varepsilon}\left(  \mu
r_{0}+1\right)  dr\rightarrow0\qquad\mathrm{when}\text{ }\varepsilon
\rightarrow0.
\end{equation}
Note that we might have chosen the following form%
\begin{equation}
\omega\left(  r\right)  =\frac{\exp\left(  -\mu r\right)  }{\mu r},
\label{omega2}%
\end{equation}
instead of $\left(  \ref{omega1}\right)  $. However, while the properties at
infinity are equal to the profile $\left(  \ref{omega1}\right)  $, close to
the throat one finds%
\begin{equation}
\omega\left(  r_{0}\right)  =\frac{\exp\left(  -\mu r_{0}\right)  }{\mu r_{0}%
}.
\end{equation}
As a consequence, the shape function becomes%
\begin{equation}
b\left(  r\right)  =r_{0}\mathit{\exp}\left(  \exp\left(  \mu r_{0}\right)
-\exp\left(  \mu r\right)  \right)
\end{equation}
while the energy density is%
\begin{equation}
\rho(r)=-\frac{\mu r_{0}}{\kappa r^{2}}\,{\mathrm{\exp}}\left(  \mu
r+\exp\left(  \mu r_{0}\right)  -\exp\left(  \mu r\right)  \right)  .
\end{equation}
If we compare the energy density obtained with $\left(  \ref{omega1}\right)  $
with the one computed with $\left(  \ref{omega2}\right)  $, we can see that,
on the throat, the amount of negative energy density is larger for the choice
$\left(  \ref{omega2}\right)  $ because%
\begin{equation}
\rho(r)=-\frac{\mu}{\kappa r_{0}}\,{\mathrm{\exp}}\left(  \mu r_{0}\right)  .
\end{equation}
This implies that the choice $\left(  \ref{omega1}\right)  $ is favored from
an energetically point of view. The same conclusion can be reached comparing
the energy density obtained with $\left(  \ref{omega2}\right)  $ with the one
obtained with $\left(  \ref{omega}\right)  $. On the other hand, when we
compare the energy density obtained with $\left(  \ref{omega}\right)  $ and
the one obtained with $\left(  \ref{omega1}\right)  $, we find%
\begin{equation}
\frac{\rho_{\omega_{2}}(r)}{\rho_{\omega_{1}}(r)}=\frac{\exp\left[
\mu\,\left(  r-r_{0}\right)  +1-{\mathrm{\exp}}\left(  \mu\,\left(
r-r_{0}\right)  \right)  \right]  }{\exp\left[  -\mu\left(  r-r_{0}\right)
\right]  }\underset{r\rightarrow\infty}{\longrightarrow}0, \label{ratio}%
\end{equation}
where $\rho_{\omega_{1}}(r)$ is the energy density obtained with the help of
$\left(  \ref{omega}\right)  $ and represented by the first component of the
SET $\left(  \ref{SET1}\right)  $, while $\rho_{\omega_{2}}(r)$ is the energy
density obtained with the help of $\left(  \ref{omega1}\right)  $ and
represented by the first component of the SET $\left(  \ref{SET2}\right)  $.
The behavior shown in $\left(  \ref{ratio}\right)  $ reveals that the
negativity of the energy density is really concentrated in the proximity of
the throat. A further investigation inspired by Yukawa profiles could be
related to the Self Sustained Traversable Wormholes, i.e., traversable
wormholes which are sustained by their own quantum
fluctuations\cite{Remo,Remo1,RGFL,RGFSNL}.

\end{document}